\documentclass{article}

\usepackage{arxiv}
\usepackage{amsthm}
\usepackage[utf8]{inputenc} 
\usepackage[T1]{fontenc}    
\usepackage{hyperref}       
\usepackage{url}            
\usepackage{booktabs}       
\usepackage{subfigure}
\usepackage{caption}
\usepackage{amssymb}
\usepackage{amsbsy}
\usepackage{amsfonts}       
\usepackage{nicefrac}       
\usepackage{color, colortbl}
\usepackage{microtype}      
\usepackage{lipsum}
\usepackage{amsmath}
\usepackage[utf8]{inputenc}
\usepackage{graphicx}
\usepackage{listings}
\usepackage{xcolor}
 
\lstdefinestyle{mystyle}{
    backgroundcolor=\color{backcolour},   
    commentstyle=\color{codegreen},
    keywordstyle=\color{magenta},
    numberstyle=\tiny\color{codegray},
    stringstyle=\color{codepurple},
    basicstyle=\ttfamily\footnotesize,
    breakatwhitespace=false,         
    breaklines=true,                 
    captionpos=b,                    
    keepspaces=true,                 
    numbers=left,                    
    numbersep=5pt,                  
    showspaces=false,                
    showstringspaces=false,
    showtabs=false,                  
    tabsize=2
}
 
\colorlet{punct}{red!60!black}
\definecolor{background}{HTML}{EEEEEE}
\definecolor{delim}{RGB}{20,105,176}
\colorlet{numb}{black}
 
\lstdefinelanguage{json}{
    basicstyle=\normalfont\ttfamily,
    numbers=left,
    numberstyle=\scriptsize,
    stepnumber=1,
    numbersep=8pt,
    showstringspaces=false,
    breaklines=true,
    frame=lines,
    backgroundcolor=\color{background},
    literate=
     *{0}{{{\color{numb}0}}}{1}
      {1}{{{\color{numb}1}}}{1}
      {2}{{{\color{numb}2}}}{1}
      {3}{{{\color{numb}3}}}{1}
      {4}{{{\color{numb}4}}}{1}
      {5}{{{\color{numb}5}}}{1}
      {6}{{{\color{numb}6}}}{1}
      {7}{{{\color{numb}7}}}{1}
      {8}{{{\color{numb}8}}}{1}
      {9}{{{\color{numb}9}}}{1}
      {:}{{{\color{punct}{:}}}}{1}
      {,}{{{\color{punct}{,}}}}{1}
      {\{}{{{\color{delim}{\{}}}}{1}
      {\}}{{{\color{delim}{\}}}}}{1}
      {[}{{{\color{delim}{[}}}}{1}
      {]}{{{\color{delim}{]}}}}{1},
}

\title{Explaining the difference between \\men's and women's football}

\author{Luca Pappalardo\\
ISTI-CNR, Italy \\
luca.pappalardo@isti.cnr.it
\And 
Alessio Rossi\\
University of Pisa, Italy \\
alessio.rossi2@gmail.com 
\And Giuseppe Pontillo\\
University of Turin, Italy \\
g.pontillo94@yahoo.it
\And Michela Natilli\\
University of Pisa, Italy\\
michela.natilli@gmail.com 
\And Paolo Cintia\\
University of Pisa, Italy\\
paolo.cintia@gmail.com}

\begin{document}
\maketitle

\begin{abstract} 
Women's football is gaining supporters and practitioners worldwide, raising questions about what the differences are with men's football.
While the two sports are often compared based on the players' physical attributes, we analyze the spatio-temporal events during matches in the last World Cups to compare male and female teams based on their technical performance. 
We train an artificial intelligence model to recognize if a team is male or female based on variables that describe a match's playing intensity, accuracy, and performance quality. 
Our model accurately distinguishes between men's and women's football, revealing crucial technical differences, which we investigate through the extraction of explanations from the classifier's decisions. 
The differences between men's and women's football are rooted in play accuracy, the recovery time of ball possession, and the players' performance quality. 
Our methodology may help journalists and fans understand what makes women's football a distinct sport and coaches design tactics tailored to female teams.   
\end{abstract}


\section{Introduction}
Women's football took its first steps thanks to the independent women of the \textit{Kerr Ladies} team, who gave the most significant impetus to this sport since the early twentieth-century \cite{scardicchio}.
As time passed, the \textit{Kerr Ladies} intrigued the English crowds for their ability to stand up to male teams in numerous charity competitions. 
The success and enthusiasm of these events aroused concerns within the English Football Association, which on December 5, 1921, decreed that ``football is quite unsuitable for females and ought not to be encouraged'', and requested ``the clubs belonging to the Association to refuse the use of their grounds for such matches'' \cite{scardicchio}. 
Unfortunately, this measure drastically slowed down the development of women's football, which, after a long period of stagnation, resurfaced in the first half of the 1960s in Europe's Nordic countries, such as Norway, Sweden, and Germany. 
From that moment on, the development of women's football was unstoppable, spreading to the stadiums of Europe and the world and carving out a notable showcase among the most popular sports in the world. 
From 2012 the number of women academies has doubled \cite{lange}, with around 40 million girls and women playing football worldwide nowadays \cite{pedersen}. 

In the last decade, the attention around women’s football has stimulated the birth of statistical comparisons with men's football \cite{lange, sakamoto, gioldasis}. 
Bradley et al. \cite{bradley} compare 52 men and 59 women, drawn during a Champions League season, and observe that women cover more distance than men at lower speeds, especially in the final minutes of the first half. 
However, at higher speed levels, men have better performances throughout the game \cite{bradley}. 
Sakamoto et al. \cite{sakamoto} examine the shooting performance of 17 men and 17 women belonging to a university league, finding that women have lower average values than men on ball speed, foot speed, and ball-to-foot velocity ratio \cite{sakamoto}. 
Pedersen et al. \cite{pedersen} question the rules and regulations of the game and, taking into account the average height difference between 20-25 years-old men and women, estimate that the ``fair'' goal height in women's football should be 2.25 m, instead of 2.44 m. 
Gioldasis et al. \cite{gioldasis} recruit 37 male and 27 female players from an amateur youth league and find that, while among male players, there is a significant difference between roles for almost all technical skills, among female players just the dribbling ability presents a significant difference. 
Sakellaris \cite{sakellaris} finds that, in international football competitions, female teams have a higher average number of goals scored per match than their male counterparts. Finally, Lange et al. \cite{lange} follow 157 female and 207 male young Dutch footballers to investigate the tendency to stop the game to permit a teammate's or opponent's care on the ground, finding that women show, on average, a greater willingness to help. 

An overview of the state of the art cannot avoid noticing that current studies focus on physical features and analyze small samples of male and female players using data collected on purpose. 
At the same time, although massive digital data about the technical behavior of players are nowadays available at an unprecedented scale and detail \cite{pappalardo2019public, pappalardo2017quantifying, gudmundsson2017spatiotemporal, decroos2019actions, bornn2018soccer, rossi2018effective}, investigations of the differences between women's and men's football from a technical point of view are still limited. 
Is the intensity of play in women's matches higher than men's ones? 
Are women more accurate than men in passing?
Furthermore, does the statistical distribution of male players' performance quality differ from that of female players?

In this article, we analyze a large dataset describing 173k spatio-temporal events that occur during the last men's and women's World Cups: 64 and 44 matches, respectively, and 32 men's and 24 women's teams with 736 male players and 546 female players. 
To the best of our knowledge, ours is the largest sample of men's and women's football matches and players.
We quantify players' and teams' performance in several ways, from the number of game events generated during a match to the proportion of accurate passes, the velocity of the game, the quality of individual performance, and teams' collective behavior.
We then tackle the following interesting question: \emph{Can a machine distinguish a male team from a female based on their technical performance only?} 

Based on the use of a machine learning classifier, we show that men's and women's football \emph{do} have apparent differences, which we investigate through the extraction of global and local explanations from the classifier's decisions. 
Opening the classifier's black box allows us to reveal that, while the intensity of the game is similar, the differences between men's and women's football are rooted in play accuracy, time to recover ball possession, and the typical performance quality of the players. 

Our methodology is useful to several actors in the sports industry.
On the one hand, a deeper understanding of female and male performance differences may help coaches and athletic trainers design training sessions, strategies, and tactics tailored for women players.
On the other hand, our results may help sports journalists tell and football fans understand what makes women's football a distinct sport.

\section{Football Data}

\begin{figure}
    \centering
    \captionsetup{type=figure}
    \subfigure[]{
    \includegraphics[width=0.45\textwidth]{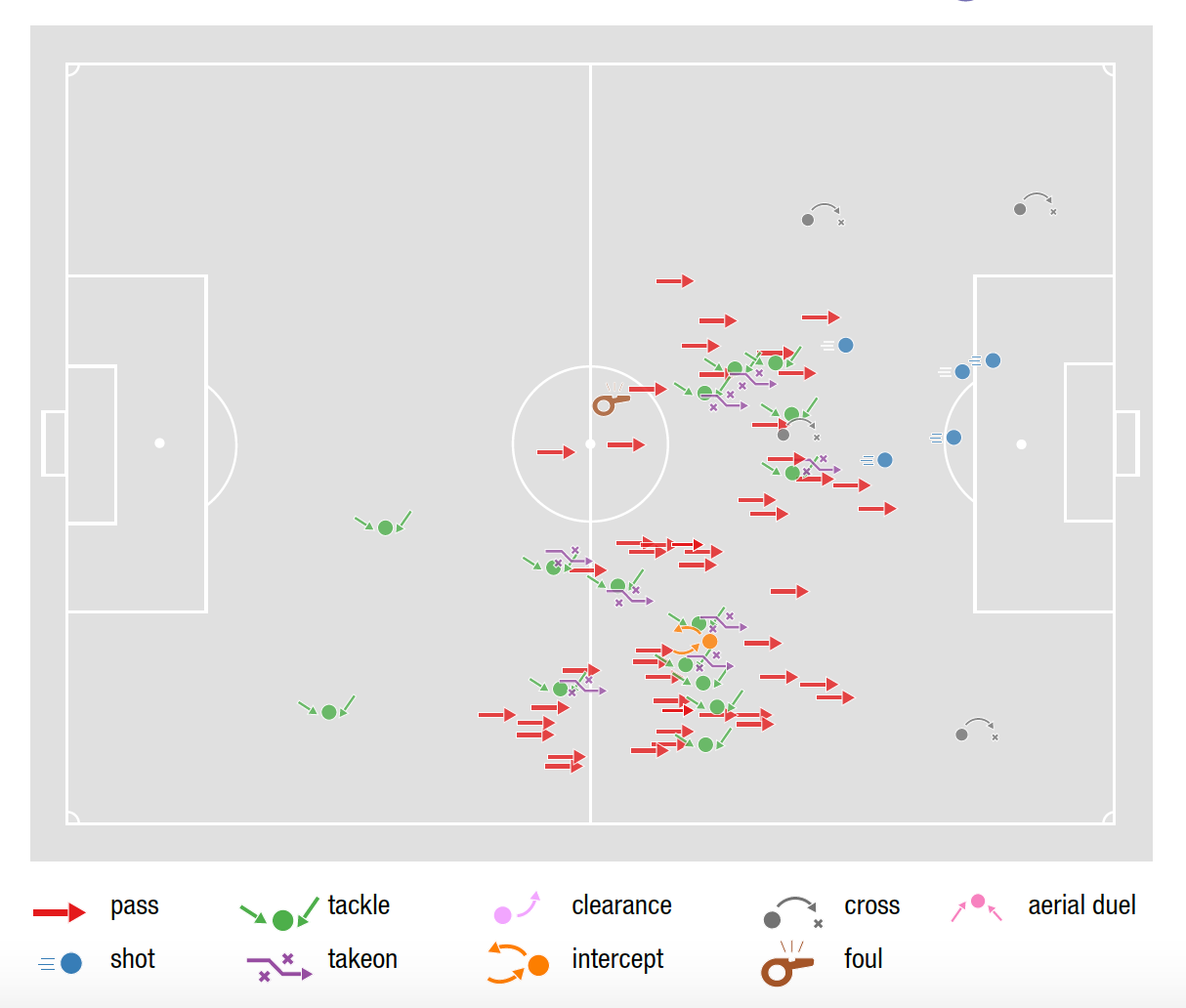}}
    \subfigure[]{
    \includegraphics[width=0.375\textwidth]{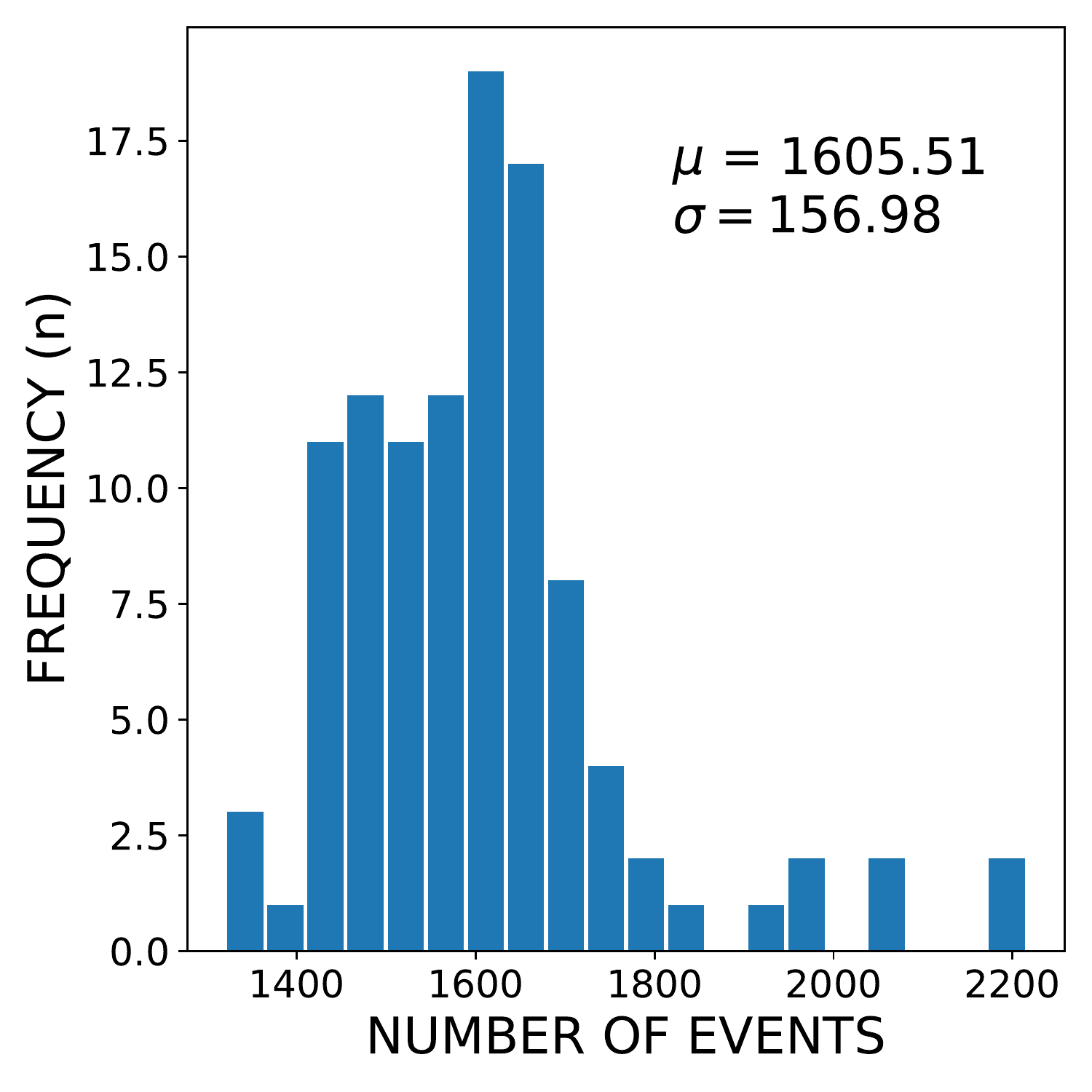}}
    \caption{(a) Example of events observed for a player in our dataset. Events are shown at the position where they have occurred. 
    This plain ``geo-referenced'' visualization of events allow understanding how to reconstruct the player's behavior during the match(b) Distribution of the number of events per match. On average, a football match in our dataset has 1600 events.}
    \label{fig:my_label}
\end{figure}

We use data related to the last men's World Cup 2018, describing 101,759 events from 64 matches, 32 national teams and 736 players, and the last women’s World Cup 2019, with 71,636 events from 44 matches, 24 national teams and 546 players. 
Each event records its type (e.g., pass, shot, foul), a time-stamp, the player(s) related to the event, the event's match, and the position on the field, the event subtype and a list of tags, which enrich the event with additional information \cite{pappalardo2019public} (see an example of event in Table \ref{fig:example_event}). 
Events are annotated manually from each match's video stream using proprietary software (the tagger) by three operators, one operator per team and one operator acting as responsible supervisor of the output of the whole match. 
The dataset regarding the men's World Cup 2018 have been publicly released recently \cite{pappalardo_massucco_2019}, in companion with a detailed description of the data format, the data collection procedure, and its reliability \cite{pappalardo2019public, playerank}.
Match event streams are nowadays a standard data format widely used in sports analytics for performance evaluation \cite{playerank, decroos2019actions, pappalardo2017quantifying, decroos2020vaep} and advanced tactical analysis \cite{decroos2018automatic, Hind, gyarmati2016competition}.
Figure \ref{fig:my_label}a shows some events generated by a player in a match.
Figure \ref{fig:my_label}b shows the distribution of the total number of events in our dataset: on average, a football match has around 1600 events, whereas a couple of matches have up to 2200 events.

\begin{figure}[htb]
\centering
\begin{lstlisting}[language=json, numbers=none]
 {"eventName": "Pass",
 "eventSec": 2.41,
 "playerId": 3344,
 "matchId": 2576335,
 "teamId": 3161,
 "positions": [{"x": 49, "y": 50}],
 "subEventName": "Simple pass",
 "tags": [{"id": 1801}]}
 \end{lstlisting}
      \captionof{table}{Example of event corresponding to an accurate pass.
      \texttt{eventName} indicates the name of the event’s type: there are seven types of events (pass, foul, shot, duel, free kick, offside and touch).
      \texttt{eventSec} is the time when the event occurs (in seconds since the beginning of the current half of the match);
      \texttt{playerId} is the identifier of the player who generated the event.
      \texttt{matchId} is the match's identifier.
      \texttt{teamId} is team's identifier.
      \texttt{subEventName} indicates the name of the subevent's type.
      \texttt{positions} is the event's origin and destination positions. 
      Each position is a pair of coordinates (x, y) in the range $[0, 100]$, indicating the percentage of the field from the perspective of the attacking team. 
      \texttt{tags} is a list of event tags, each describing additional information about the event (e.g., accurate).
      A thorough description of this data format and its collection procedure can be found in \cite{pappalardo2019public}.
      }
      \label{fig:example_event}
\end{figure}
  

\section{Technical Performance}
\emph{Do technical characteristics of men's and women's football significantly differ, statistically speaking?} 
To answer this question, we define variables that describe relevant technical aspects of the game and show for which of them there is statistical difference between men and women. 
In particular, we investigate three technical aspects: \emph{(i)} intensity of play (Section \ref{sec:intensity}); \emph{(ii)} shooting distance (Section \ref{sec:shooting_distance}); and \emph{(iii)} performance quality (Section \ref{sec:performance}). 

\subsection{Intensity of play} 
\label{sec:intensity}
The intensity of play is associated with a team's chance of success \cite{Hind, cintia2015network}.
Here, we measure intensity of play in terms of volume and velocity.

\paragraph{Volume.}
For each team in a match, we compute the total number of events and the number of specific event types (duels, fouls, free kicks, offsides, passes and shots) \cite{pappalardo2019public}.
Although, on average, men’s matches show more events that women’s ones, this difference is not statistically significant (unpaired t-score = 1.40, p-value = 0.16, see Table \ref{eventstable}).
Women's matches have, on average, more free kicks, duels, others on the ball (i.e., accelerations, clearances and ball touches) and passes but fewer fouls than men's matches (Table \ref{eventstable}). Additionally, men's passes are also on average more accurate than women's ones (unpaired t-score = 8.95, p-value $<$ 0.001).


\paragraph{Velocity.}
The average pass velocity $\mbox{PassV}(g)$ measures the average time between two consecutive passes in a match $g$, and the average ball recovery time $\mbox{RecT}(g)$ measures the average time for a team to recover ball possession in $g$ (see Supplementary Information 1). 
The interruption time $\mbox{StopT}(g)$ indicates the time spent between two consecutive actions (i.e., time to make a free-kick, a corner kick or a throw-in).
The average pass length $\mbox{PassL}(g)$ measures the average time between a team's two consecutive shots in a match and the average distance between a pass's starting and ending points, respectively.
For all of these features, we perform
an unpaired t-test to detect differences between men and women (Table \ref{eventstable}).
We find that 
women's $\mbox{PassV}(g)$  (unpaired t-score = 8.69, p-value $<$ 0.001) is lower than men's one, denoting a higher velocity of passes in men's football (unpaired t-score = 3.540, p-value $<$ 0.001). At the same time, 
women's $\mbox{RecT}(g)$ is lower than male's one (unpaired t-score = 5.41, p-value $<$ 0.001), i.e., women regain ball possession faster. 
In contrast, 
men's passes $\mbox{PassL}(g)$ (unpaired t-score = 3.54, p-value $<$ 0.001) are on average larger than women's ones.


\definecolor{Gray}{gray}{0.9}
\begin{table}\centering
\begin{tabular}{l|c|c|c|c}
\toprule
\textbf{Event} & \textbf{Women} & \textbf{Men} & \textbf{t-score} & \textbf{p-value}\\ 
\midrule

\# events & 1522.62$\pm$93.82 & 1549.62$\pm$99.55 & 1.40 & 0.16 \\ 
\hline

\# shots & 21.98$\pm$6.03  & 21.52$\pm$5.72 & -0.40 & 0.69 \\ 
\hline

\rowcolor{Gray}
\# fouls & 19.95$\pm$5.94 & \bf 26.94$\pm$6.41 & 5.68 & $<$0.001 \\ 
\hline


\rowcolor{Gray}
\# passes & \bf 861.67$\pm$101.25 & 790.86$\pm$98.76 & 3.57 & 0.001 \\ 
\hline

\rowcolor{Gray}
\# free kicks & \bf 102.70$\pm$11.85 & 90.05$\pm$10.62 & -5.75 & $<$0.001 \\ \hline

\rowcolor{Gray}
\# duels & \bf 419.91$\pm$53.77 & 394.52$\pm$62.25 & -2.18 & 0.03 \\ \hline

\rowcolor{Gray}
\# offside & \bf 3.88$\pm$2.91 & 2.91$\pm$1.86 & -2.39 & 0.02 \\ 
\hline

\rowcolor{Gray}
\# others & \bf 149.98$\pm$26.08 & 141.19$\pm$24.65 & -1.76 & 0.05 \\ \hline

\rowcolor{Gray}
\# accurate passes & 311.66$\pm$127.17 & \bf 375.67$\pm$138.30 & 3.49 & $<$0.001 \\ \hline

\rowcolor{Gray}
Pass accuracy (AccP) & 0.76$\pm$0.08 & \bf 0.84$\pm$0.05 & 8.95 & $<$0.001 \\ \hline

\rowcolor{Gray}
Pass velocity (PassV) & 2.83$\pm$0.12 & \bf 2.99$\pm$0.17 & 8.69 & $<$0.001 \\ 
\hline

\rowcolor{Gray}
Recovery Time (RecT) & 19.58$\pm$10.37 & \bf 27.32$\pm$10.14 & 5.41 & $<$0.001 \\ 
\hline

\rowcolor{Gray}
Stop time (StopT) & 18.92$\pm$3.38 & \bf 23.27$\pm$2.99 & 6.98 & $<$0.001 \\ 
\hline

\rowcolor{Gray}
Pass lenght (PassL) & 19.53$\pm$1.53 & \bf 20.32$\pm$1.70 & 3.54 & $<$0.001 \\ 
\hline

\rowcolor{Gray}
Shot distance (ShotD) & 18.39$\pm$1.90 & \bf 19.99$\pm$1.74 & 4.47 & $<$0.001 \\ 
\hline

\rowcolor{Gray}
Average PR (PR$_{avg}$) & -0.01$\pm$0.01 & \bf 0.01$\pm$0.01 & 9.01 & $<$0.001 \\ 
\hline

Standard deviation PR (PR$_{std}$) & 0.05$\pm$0.03 & 0.05$\pm$0.03 & -0.40 & 0.69 \\ 
\hline

\rowcolor{Gray}
H-indicator (H) & 1.21$\pm$0.27 & \bf 1.32$\pm$0.36 & 2.49 & 0.01 \\
\hline

\rowcolor{Gray}
Flow centrality (FC)  & 0.058$\pm$0.004 & \bf 0.059$\pm$0.003 & 2.11 & 0.04 \\
\bottomrule
\end{tabular}
\caption{Statistical difference of technical features between male and female teams. 
The summary data for both women and men are report as mean$\pm$standard deviation per matches. 
Grey rows indicates features for which the difference between men and women is statistically significant. The highest values are highlighted in bold.}
\label{eventstable}
\end{table}

\subsection{Shooting distance} 
\label{sec:shooting_distance}

We explore the spatial distribution of the positions where male and female players perform free kicks and shots (see Supplementary Figure \ref{shots&kicks}) and quantify shooting distance ShotD as the Euclidean distance from the position where the shots starts to the center of the opponents' goal. 
To find statistical difference between men and women, we use the non-parametric Mann-Whitney U-Test. 
On average, men players kick the ball from a greater distance than women (p-value < 0.001, Table \ref{eventstable}).

To take into account that men and women may have a different perception of distance to the opponents' goal, we split the attacking midfield into three zones $Z1$, $Z2$ and $Z3$, according to the two distributions of shooting distance, i.e., looking at a shot's minimum and the maximum starting positions. 
$Z1$ is the area closest to the goal, $Z3$ the furthest, $Z2$ the zone in the middle. 
The zones of women are 1.1 meters closer to the goal than the zones of men (p-value < 0.001).

We then use a z-test for proportions with two independent samples to verify whether there is a difference in the shooting activity between men and women. 
Female teams have a higher percentage of shots from their Z1 zone than male teams (p-value = 0.01); the opposite is true in the Z2 shooting area (p-value = 0.004). 
Finally, female teams have a higher percentage of shots from their Z3 shooting area (p-value = 0.02) than male teams.


\subsection{Performance quality}
\label{sec:performance}
\begin{figure} 
\centering
\includegraphics[width=0.9\textwidth]{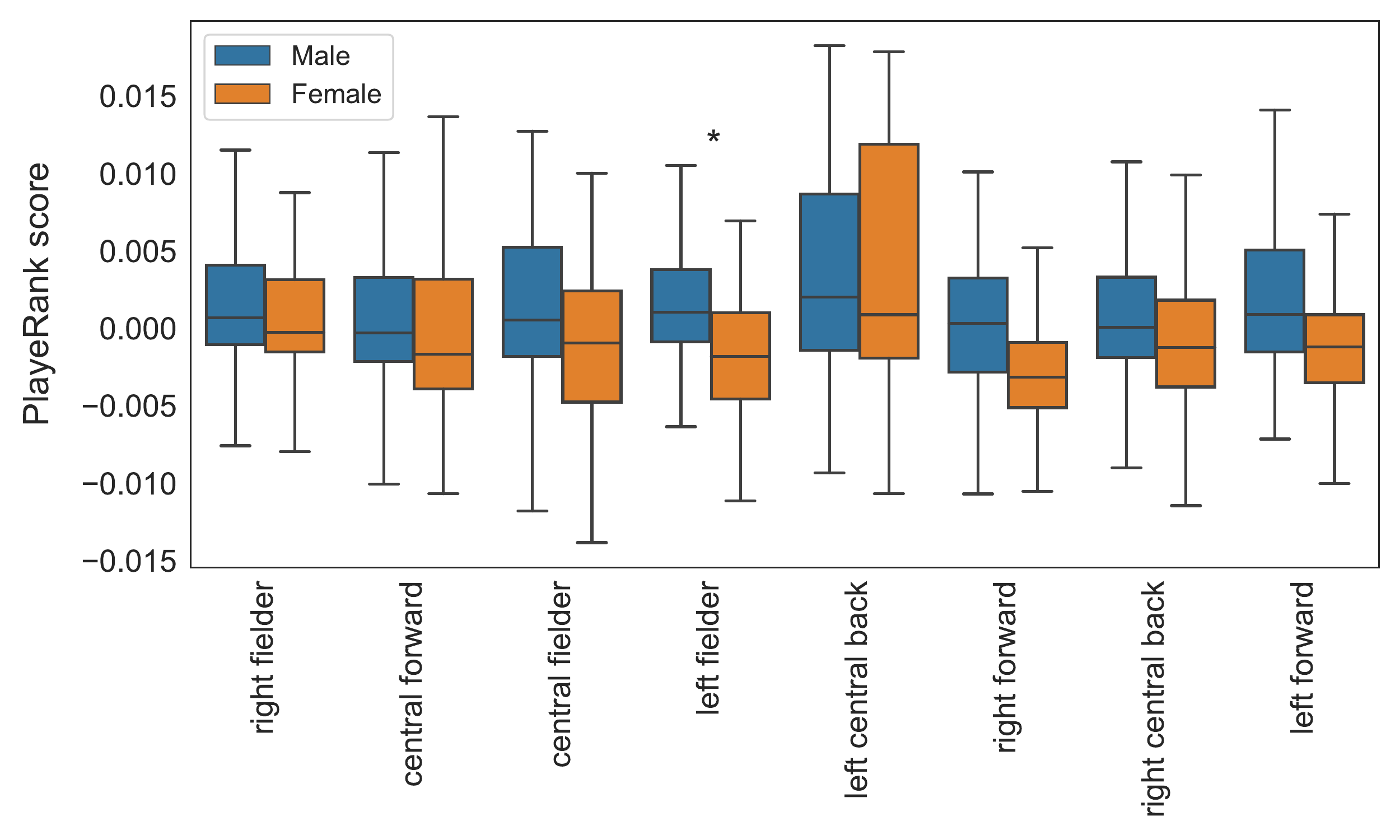}
\caption{PlayeRank score by role fro male and female players. Asterisks indicate significant statistical difference between male and female for that role.}
\label{fig:PR}
\end{figure}

We use the PlayeRank algorithm \cite{playerank} to compute the PR score, which quantifies a player's performance quality in a match (see Supplementary Information 2 for details on the algorithm).
PlayeRank is robust in agreeing with a ranking of players given by professional football scouts, given its capability of describing football performance comprehensively \cite{playerank}.
For each match $g$, and for both teams, we compute the mean and the standard deviation of the individual PR scores, $\mbox{PR}_{avg}(T, g)$ and $\mbox{PR}_{std}(T, g)$, respectively.
High values of $\mbox{PR}_{avg}(T, g)$ indicate that the players in team $T$ perform well in match $g$, on average.
High values of $\mbox{PR}_{std}(T, g)$ indicate a large variability of PR across the teammates in match $g$.
Male players have higher PR$_{avg}$ than females players (unpaired t-score = 9.01, p<0.001) but similar PR$_{std}$ (unpaired t-score = -0.40, p-value = 0.69). We find statistical difference in the PR score between men and women for left fielders only (Figure \ref{fig:PR}).

We also explore the differences in the collective behavior of male and female teams computing the passing networks, graphs in which nodes are players and edges represent passes between teammates in a match \cite{Hind, amaral2010quantifying, pena2012network, cintia2016haka, buldu2019defining}. 
From the passing network of a team $T$ in a match $g$ we derive the H indicator $H(T, g)$ \cite{Hind,cintia2015network} and the team flow centrality $\mbox{FC}(T, g)$ \cite{amaral2010quantifying}, two ways of quantifying the goodness of a team's performance in a match \cite{pappalardo2019public} (Supplementary Information 3). 
$H(T, g)$ summarizes different aspects of a team's passing behaviour, such as the average amount $\mu_p$ of passes and the variance $\sigma_p$ of the number of passes managed by players \cite{Hind}.
The higher the $\sigma_p$, the higher is the heterogeneity in the volume of passes managed by the players. 
A player's flow centrality in a match is defined as their betweenness centrality in the passing network \cite{amaral2010quantifying}.
The team flow centrality, $\mbox{FC}(T, g)$, is hence defined as the average of the flow centralities of players of team $T$ in match $g$ \cite{amaral2010quantifying}.

Table \ref{tab_H_PR_FC} shows the top ten male and female teams with highest average H indicator H$_{avg}$, the average PR score $\mbox{PR}_{avg}(T)$, and average FC score $\mbox{FC}_{avg}(T)$. 
Spain is the male team with the best overall team performance ($H_{avg}^{(M)}(\mbox{\footnotesize Spain}) = 1.67$), and so is Japan in the women's World Cup ($H_{avg}^{(F)}(\mbox{\footnotesize Japan}) = 1.56$). 
In general, the H indicator of male teams ($H_{avg}^{(M)} = 1.32$) is higher (unpaired t-score = 2.67, p<0.02) than female teams' one ($H_{avg}^{(F)} = 1.21$).
Similarly, the FC indicator of male teams ($FC_{avg}^{(M)} = 0.059$) is slightly higher (unpaired t-score = 2.11, p<0.04) than female teams' one ($FC_{avg}^{(W)} = 0.058$).

\begin{table}\centering
\caption{List of the top ten football teams with the highest average H, FC, and PR indicators in the two competitions.}
\begin{tabular}{|l|c|r|r|r|}
\hline
\textbf{Team} & \bf Sex & $H_{avg}$ \\ \hline
Spain & M & 1.67 \\ \hline
Egypt & M & 1.60 \\ \hline
Denmark & M & 1.59 \\ \hline
Japan & F & 1.56 \\ \hline
Australia & M & 1.54 \\ \hline
England & F & 1.53 \\ \hline
Chile & F & 1.51 \\ \hline
Iran & M & 1.47 \\ \hline
England & M & 1.45 \\ \hline
Tunisia & M & 1.44 \\ \hline
\end{tabular}
\quad
\begin{tabular}{|l|l|l|}
\hline
\textbf{Team} & \textbf{Sex} & $PR_{avg}$ \\ \hline
USA & F & 0.08 \\ \hline
France & F & 0.04 \\ \hline
Belgium & M & 0.04 \\ \hline
Germany & F & 0.04 \\ \hline
Australia & F & 0.035 \\ \hline
Italy & F & 0.035 \\ \hline
Croatia & M & 0.035 \\ \hline
Sweden & F & 0.034 \\ \hline
Russia & M & 0.03 \\ \hline
England & F & 0.03 \\ \hline
\end{tabular}
\quad
\begin{tabular}{|l|l|l|}
\hline
\textbf{Team} & \textbf{Sex} & $FC_{avg}$ \\ \hline
Mexico & M & 0.064 \\ \hline
Germany & M & 0.063 \\ \hline
Morocco & M & 0.063 \\ \hline
Spain & M & 0.063 \\ \hline
Argentina & M & 0.062 \\ \hline
USA & F & 0.062 \\ \hline
Canada & F & 0.061 \\ \hline
Japan & F & 0.061 \\ \hline
England & F & 0.061 \\ \hline
Peru & M & 0.06 \\ \hline
\end{tabular}
\label{tab_H_PR_FC}
\end{table}

\subsection{In Summary}
Our statistical analysis reveals that male and female teams \textbf{do} differ in many technical characteristics (Table \ref{eventstable}):
\begin{itemize}
\item Men perform more passes per match with a higher accuracy indicating a higher volume of play and a better technical quality of the men compared to woman;
\item Men perform longer passes and shoot from a longer distance than women, presumably due to the physical differences between genders (e.g., men have greater strength in the legs, which allows them to shoot from farther away);
\item The typical performance quality of male teams, in terms of pass volume, heterogeneity, centrality and PR score, is higher than women's one. This result could be related to the different player style; 
\item 
Women's ball recovery time is shorter than men's, denoting either a better capability of women to recover ball or a lower capability to retain it, and characterizing a more fragmented game in women's football. 
\end{itemize}

\section{Team gender recognition}
\label{sec:gender}

Having established that women's and men's football differ in many technical characteristics related to intensity of play, shooting distance, and performance quality, we now turn to the question: \emph{Can we design a machine learning classifier to distinguish between a male and a female football team?}  
Machine learning can capture the interplay between technical features, and explanations extracted from the constructed classifier can reveal further insights on the differences between men and women football \cite{guidotti2018survey}.

As a first step, we describe the behavior of a team $T$ in match $g$ by a performance vector of $19$ variables and associate it with a target variable:

\begin{itemize} 
\item number of events (\# events) and number of events of each type (\# shots, \# fouls, \# passes, \# free kicks, \# duels, \# offside, \# others, \# accurate passes); 
\item percentage of accurate passes $\mbox{AccP}$, average shooting length $\mbox{ShootL}$ and average pass length $\mbox{PassL}$; 
\item average time between passes $\mbox{PassV}$;
\item average time to regain ball possession $\mbox{RecT}$ and how long a team takes before re-starting the game through a free-kick, a corner kick or a throw-in $\mbox{StopT}$; 
\item the H-indicator $\mbox{H}$, the team flow centrality $\mbox{FC}$, the average PR score $\mbox{PR}_{avg}$ and its standard deviation $\mbox{PR}_{std}$. 
\item the target variable indicates whether the team is male (class 1) or female (class 0).
\end{itemize}

We build a supervised classifier and use 20\% of the dataset to tune its hyper-parameters through a grid search with 5-folds cross validation.
We use the remaining 80\% of the dataset to validate the model using a leave-one-team-out cross-validation: in turn, we leave out all matches of one team and train the model using all matches of the remaining teams. 
We assess the performance of the model using four metrics \cite{elements}: 
\emph{(i)} accuracy, the ratio of correct predictions over the total number of predictions;
\emph{(ii)} precision, the ratio of correct predictions over the number of predictions for the positive class (male);
\emph{(iii)} recall, the ratio of correct predictions over the total number of instances of the positive class (male);
\emph{(iv)} F1 score, the harmonic mean of precision and recall.


We try several learners to construct different types of classifiers (Decision Tree, Logistic Regression, Random Forest, and AdaBoost). 
All classifiers achieve a good performance (see Supplementary Figure \ref{roc}), with an average relative improvement of 67\% in terms of F1-score over a classifier that always predicts the team's gender randomly (Table \ref{clasresult}).
The best model, AdaBoost, has an improvement of 93\% over the baseline in terms of F1-score.
These results indicate that a classifier can distinguish between male and female teams on the only basis of the performance variables. 

\begin{table}
\centering
\begin{tabular}{|l|c|c|c|c|}
\hline
\textbf{Classifier} & \textbf{Accuracy} & \textbf{Precision} & \textbf{Recall} & \textbf{F1-Score} \\ \hline
AdaBoost.M1 & \bf 0.93 (93\%) & \bf 0.80 (70\%) & \bf 0.92 (119\%) & \bf 0.85 (93\%) \\ \hline
Random Forest & 0.86 (46\%) & 0.69 (45\%) & 0.82 (95\%) & 0.73 (65\%) \\ \hline
Decision Tree & 0.85 (77\%)& 0.68 (44\%)& 0.79 (88\%) & 0.71 (61\%) \\ \hline
Logistic & 0.79 (64\%) & 0.64 (36\%) & 0.79 (88\%) & 0.66 (50\%) \\ \hline 
\hline
Baseline & 0.48 & 0.47 & 0.42 & 0.44 \\ \hline
\end{tabular}
\caption{Table of the leave-one-team out cross-validation results (i.e., Accuracy, Precision, Recall and F1-score) computed on the training dataset of each machine learning classifiers used to predict a football team in a game as male (\textit{class 0}) or female (\textit{class 1}). The \textit{baseline classifier} always predicts by respecting the training set’s class distribution, which is balanced. The percentages in the table refer to the improvement of machine learning model compared to the baseline results.}
\label{clasresult}
\end{table}

The inspection of the reasoning underlying the model's decisions can provide us deeper insights into the differences between men's and women's football. 
We extract global (i.e., inference on the basis of a complete dataset) and local (i.e., inference about an individual prediction) explanations from the best model (AdaBoost) using SHAP\footnote{library released for Python (\url{https://github.com/slundberg/shap})}, a method to explain each prediction based on the
optimal Shapley value \cite{lundberg2017unified}.
The Shapley value of a performance variable is obtained by composing a combination of several variables and average change depending on the presence or absence of the variables to determine the importance of a single variable based on game theory \cite{lundberg2017unified}.
The interpretation of the shapley value for variable value $j$ is: the value of the $j$-th variable contributed  
$\phi_j$ to the prediction of a particular instance compared to the average prediction for the dataset \cite{molnar2020interpretable}.

\begin{figure}[htb] 
\centering
\includegraphics[width=0.8\textwidth]{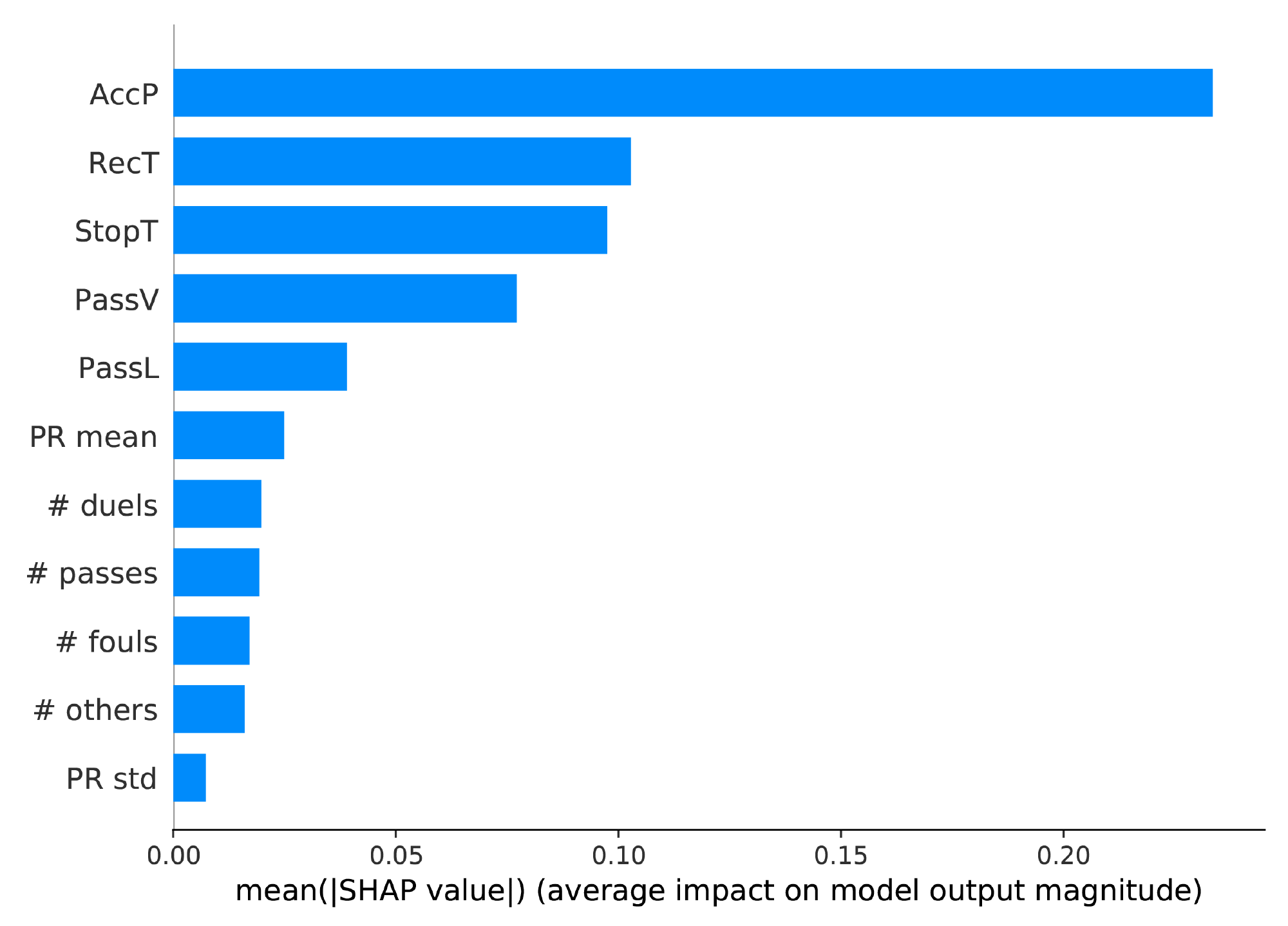}
\caption{Ranking of features importance (mean Shap value) extracted from the team gender classifier.}
\label{fig:global}
\end{figure}

Figure \ref{fig:global} shows the global explanation of AdaBoost, in which variables are ranked based on their overall importance to the model in accordance with shap values. 
Pass accuracy (AccP) is way far the most important feature to classify a team's gender. 
Recovery time (RecT), average interruption time (StopT), pass velocity (PassV), pass length (PassL), \# duels and \# passes, PR$_{avg}$, \# fouls and PR$_{std}$ are other important features for the decision making process.

\begin{figure}[htb] 
\centering
\includegraphics[width=0.8\textwidth]{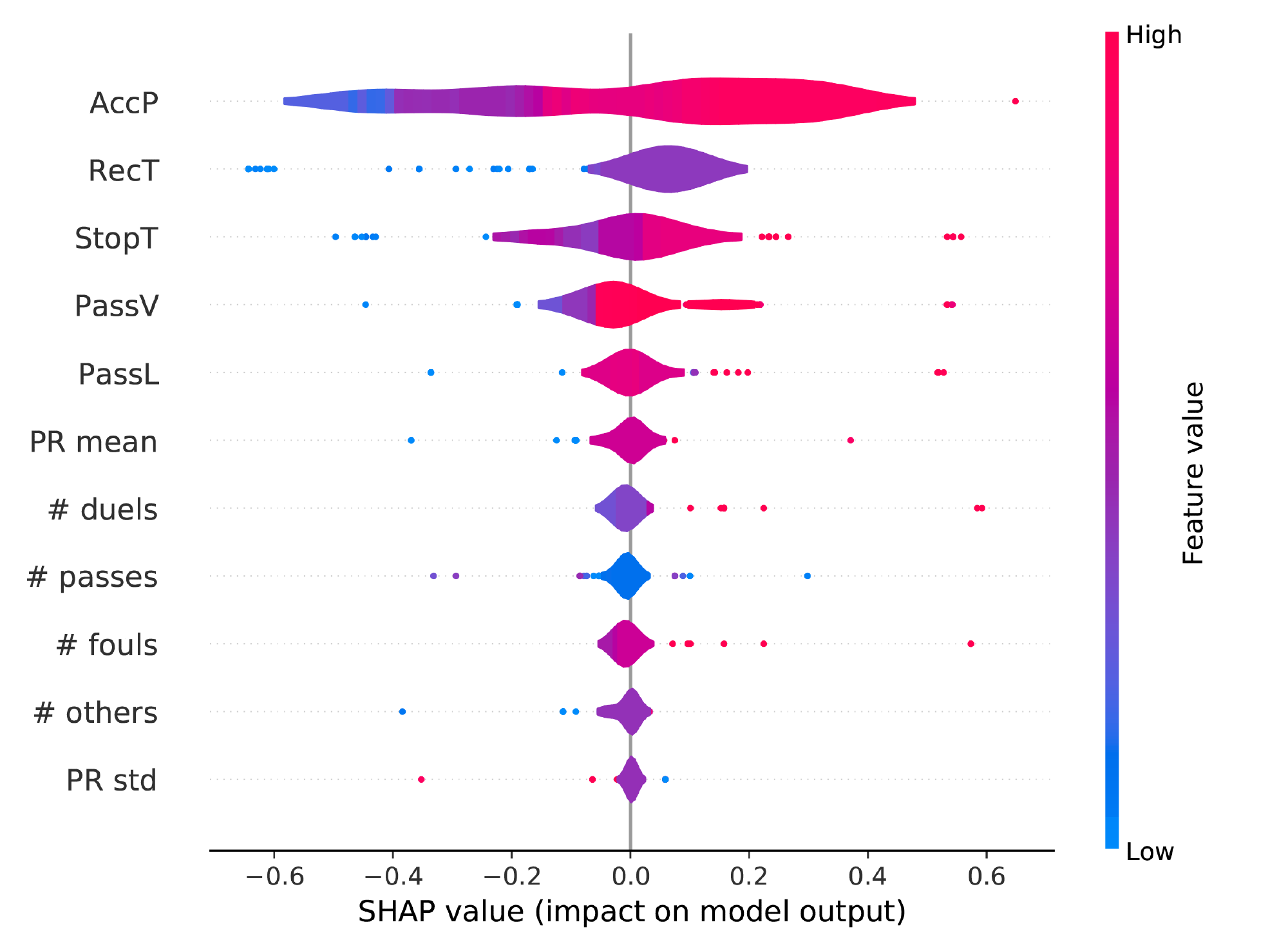}
\caption{Distribution of the impact of each feature on the team gender classifier. 
The color represents the feature value (red high, blue low); and position of the point indicates the Shap value. }
\label{fig:global_2}
\end{figure}

Figure \ref{fig:global_2} shows the summary plot that combines feature importance and feature effects, where each point indicates a team. 
The position of a feature on the y-axis indicates the importance of that feature to the model's decision. 
A point's color, in a gradient from blue (low) to red (high), indicates its numerical value.
The position of a point on the x-axis indicates the associated shap value: positive values indicate that a team is more likely to be male; negative values that it is more likely to be female.
Higher values of PassAcc (red points) are associated with higher shap values.  
This indicates that male players are typically more accurate in passing, a property that is used by the classifier to discriminate a male team from a female one.
Similarly, high values of RecT are associated with a higher probability of a team to be male, highlighting \emph{a fortiori} that female teams are characterized by a more fragmented play.

\begin{figure}[htb] 
\centering
\includegraphics[width=0.9\textwidth]{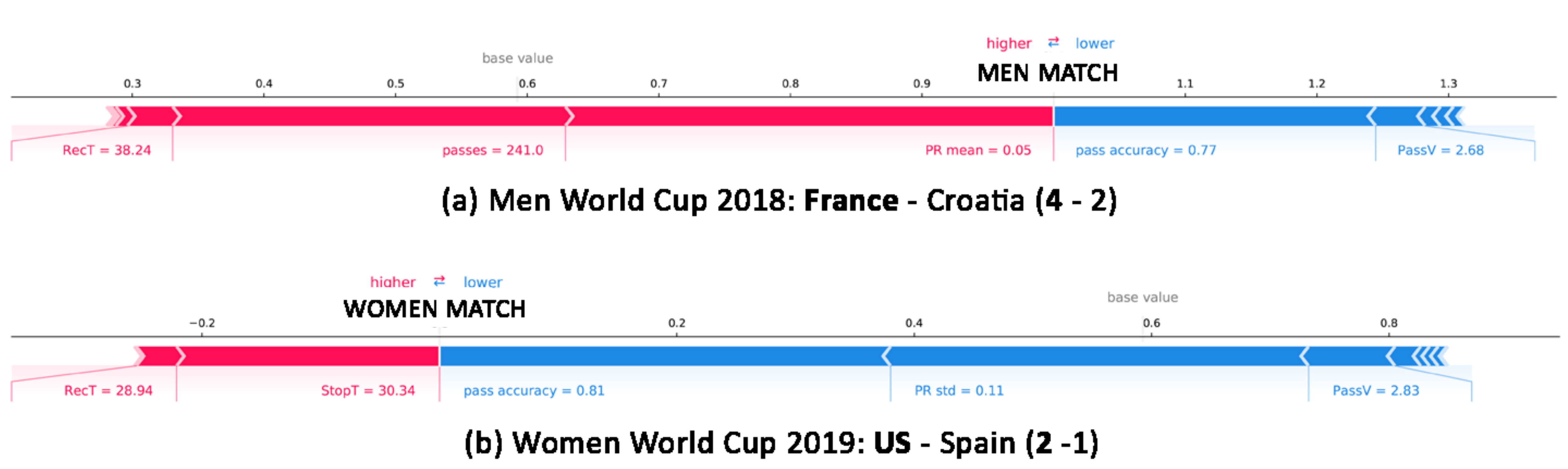}
\caption{Local Shap explanations for two examples in our dataset: France in France vs Croatia and USA in match USA vs Spain. Feature values that increase the probability of a team to be male are shown in red, those decreasing the probability are in blue.}
\label{fig:example_pred}
\end{figure}

Figure \ref{fig:example_pred}a refers to the final of the men's World cup 2018, Croatia vs. France. 
AdaBoost correctly predicts that France is a male team, basing its decision on five main variables: PR$_{avg}$, \#passes, AccP, PassV, and RecT.
France has RecT(France, \mbox{\footnotesize CRO vs FRA}) $ = 38.24$, \#passes(France, \mbox{\footnotesize CRO vs FRA}) $=241$ and PR$_{avg}$(France, \mbox{\footnotesize CRO vs FRA}) $=0.05$, closer to the typical values of men's football (RecT$^{(M)} = 27.32$, \#passes$^{(M)}= 394.43$, PR$_{avg}^{(M)} = 0.01$) than to those of female's football (RecT$^{(F)} = 19.58$, \#passes$^{(F)}= 430.84$, PR$_{avg}^{(F)} = -0.01$).
In contrast, AccP(France, \mbox{\footnotesize CRO vs FRA}) $ = 0.77$ and PassV(France, \mbox{\footnotesize CRO vs FRA}) $ = 2.68$, which are closer to the typical values of a female team (AccP$^{(F)} = 0.76$, PassV$^{(F)}=2.83$, Table \ref{eventstable}) than to those of a male one (AccP$^{(M)} = 0.84$, PassV$^{(M)}=2.99$, Table \ref{eventstable}).
Overall, the sum of the shap values indicates that France played a match in accordance with the typical characteristics of a male team. 

Figure \ref{fig:example_pred}b shows the prediction of a match in the women's World Cup 2019, USA vs Spain. 
In this case, AdaBoost correctly predicts that USA is a female team, basing its decision mainly on AccP, PR$_{std}$, StopT, RecT, and PassV.
USA has RecT(USA, \mbox{\footnotesize USA vs SPA}) $ = 28.94$ and StopT(USA, \mbox{\footnotesize USA vs SPA}) $ = 30.34$, closer to the typical values of men's football (RecT$^{(M)}=27.32$ and StopT$^{(M)}=23.27$, Table \ref{eventstable}) than to those a women's football (RecT$^{(M)}=19.58$ and StopT$^{(M)}=18.92$, Table \ref{eventstable}). 
In contrast, the values of AccP(USA, \mbox{\footnotesize USA vs SPA}) $ = 0.81$ and PassV(USA, \mbox{\footnotesize USA vs SPA}) $ = 2.83$, more similar to those of women teams (Table \ref{eventstable}). 
Overall, the sum of the shap values leads the model to classify US as a female team.


\begin{figure}[htb]
\centering
\subfigure{\includegraphics[scale=.39]{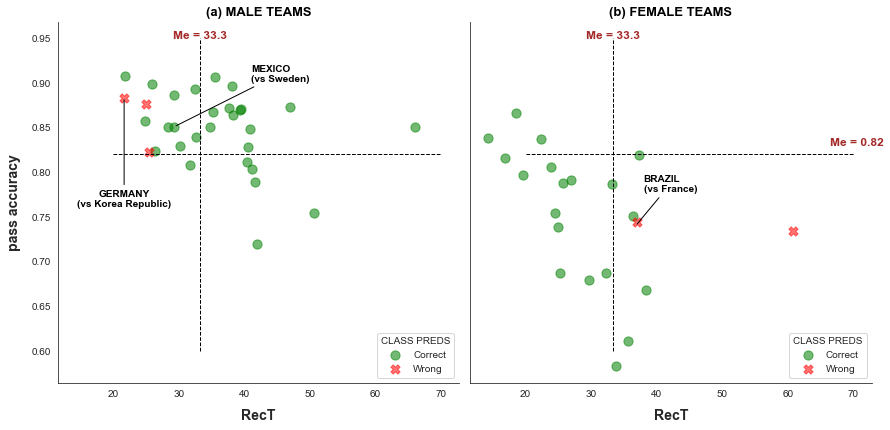}}
\subfigure{\includegraphics[width=0.9\textwidth]{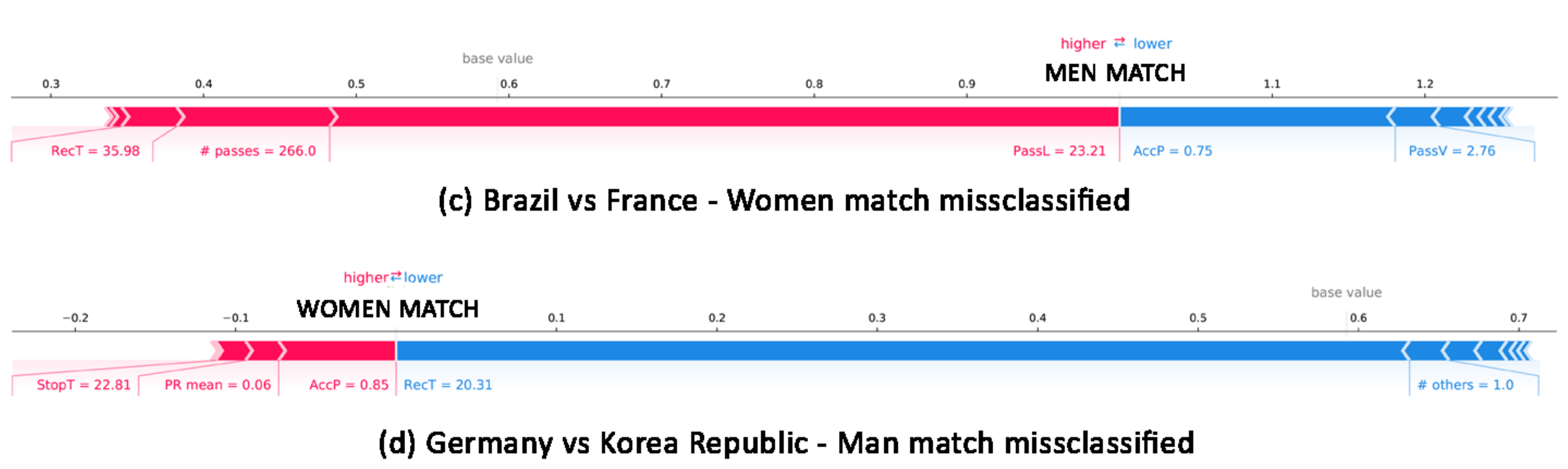}}

\caption{(a, b) Scatter plots displaying AccP versus RecT for a test set of teams in male matches (a) and female matches (b). 
Circles indicate a team correctly classified by the team gender classifier, crosses indicate a mistake by the classifier. 
The dashed lines are at the median values for the two variables over the entire data set.
In plots (c) and (d) we report the local Shap explanations of two misclassified examples.
}
\label{test_predictions} 
\end{figure}

Figure \ref{test_predictions}a and \ref{test_predictions}b visualize the predictions of the AdaBoost classifier on a test set of 31 men's matches and and 21 women's matches concerning the two most important variables, $\mbox{AccP}$ and $\mbox{RecT}$. 
In just two cases out of 21, AdaBoost misclassifies a female team as a male one (Figure \ref{test_predictions}b).
For example, in match Brazil vs France of the women's World Cup, RecT(Brazil, \mbox{\footnotesize BRA vs FRA}) $ = 35.89$ and AccP(Brazil, \mbox{\footnotesize BRA vs FRA}) $ = 0.75$ (Figure \ref{test_predictions}c), which leads the model to misclassify it as a male team because those values are more typical of women's football than of men's football.

In just three cases out of 31, a male team is misclassified as a female one (Figure \ref{test_predictions}a, red crosses). 
For example, in match Sweden vs Mexico of the men's World Cup, Mexico is correctly classified as a male team: its values of  $\mbox{AccP}(\mbox{\mbox{Mexico}, \mbox{\footnotesize SWE vs MEX}}) = 0.85$ and $\mbox{RecT}(\mbox{Mexico}, \mbox{\footnotesize SWE vs MEX})= 30$ are indeed close to the typical values of men's football.
In contrast, in match Germany vs. South Korea, Germany is misclassified as a female team, mainly because $\mbox{RecT}(\mbox{Germany}, \mbox{\footnotesize GER vs KOR})=20.31$ makes it more similar to a female team (RecT$^{(F)} = 19.58$) than to a male one (RecT$^{(M)} = 27.32$, see Table \ref{eventstable} and Figure \ref{test_predictions}d).


The misclassified women's teams have on average $\mbox{AccP}^{(F, \mbox{\scriptsize wrong})}=0.76 > \mbox{AccP}^{(F)}=0.75$, and a $\mbox{RecT}^{(F, \mbox{\scriptsize wrong})} = 31 > \mbox{RecT}^{(F)}= 29$. 
Moreover, on average $\mbox{StopT}^{(F, \mbox{\scriptsize wrong})}= 19$, which is greater than $\mbox{StopT}^{(F)}=18$ among all female teams. 
The misclassified male teams have $\mbox{AccP}^{(M, \mbox{\scriptsize wrong})} = 0.81 < \mbox{AccP}^{(M)} = 0.84$ (close to $\mbox{AccP}^{(F)} = 0.75$), and $\mbox{RecT}^{(M, \mbox{\scriptsize wrong})}= 36 < \mbox{RecT}^{(M)}=37$ ($\mbox{RecT}^{(F)}= 29$).
In both cases, AccP and RecT play a fundamental role in confusing the classifier.

\section{Conclusions}
The availability of spatio-temporal match events related to the last men's and women's World Cups allowed us to compare the technical characteristics of men's and women's football.
While most of the existing works focus on the differences in physical characteristics, we reconstructed a complex mosaic of the differences between male and female players. 
Our statistical analysis revealed that differences do exist in several technical features: the time between two consecutive events and the time required to recover possession are the lowest in women's football; conversely, male teams are typically more accurate in passing, and they kick the ball from a greater distance than women players.
The inspection, through global and local explanations, of a model that classifies team gender from the technical features, confirmed that the percentage of accurate passes and the time to recover possession are crucial to distinguish between the two sports.  
In particular, the usage of the local explanations provide a novel perspective to reason about the difference between men and women in football, highlighting the reason behind the peculiar cases in which the classifier has been ``fooled'' by a team's technical performance.

Our results are open to various interpretations.
First of all, the statistical non-significance of the difference in the number of events and shots suggest that, overall, men's and women's football have similar play intensity.
Conversely, the higher accuracy of passes in men's matches may be due to the higher technical level of male players, which may be rooted in the fact that national teams in the men's World Cup are mainly composed of professional players. 
In contrast, several female national teams (e.g., Italy) are composed of non-professional players or professional players for a short time.
Although women's football's technical level is increasing rapidly, there is still a technical gap between the two sports.
The shorter recovery time observed for women's matches may be due to both the lower pass accuracy (i.e., more balls lost) and a better capacity of women to press the opponents and recover ball possession.
Performance indicators reveal that centrality is higher in men's football, denoting the presence of ``hub'' players that centralize the game (higher flow centrality) and higher variability in the performance quality across teammates (higher H indicator and PR score).
This suggests that women's football passes are more uniformly distributed across the teammates.
Women's football also has a preference for short passes over long balls. 
Since accurate long balls are harder than short ones, this preference may be a solution to compensate for women players' lower technical level.

As future work, we plan to investigate differences in men's and women's football in national tournaments, and to investigate to what extent these differences vary nation by nation and between national and continental competitions. 
Are the difference we found in this paper more marked in the longer competitions for clubs?

\bibliographystyle{acm} 
\bibliography{references}

\appendix

\section{Supplementary Information} \label{info}

\subsection*{Supplementary Information 1: Intensity of Play} \label{info::match_intensity}

We split a match into possession phases, i.e., sequence of consecutive events in which one team only owns the ball \cite{pappalardo2019public}.
An action begins when a team gains the ball and ends if one of these cases occurs: the first half or the second half of match end, the ball goes out of the field, there is an offside or a foul \cite{pappalardo2019public}. 
In women's matches there is an event that is not present in men's matches, the so-called \textit{cooling breaks}, i.e., pauses in the game due to excessive heat; the algorithm recognizes them and indicates them as an additional cause of end of action. 

\paragraph{Average pass velocity.} The average pass velocity $\mbox{PassV}(g)$ in a match $g$ is the average time between two consecutive passes in which the receiver of the first pass is the player who makes the next pass to a teammate.

\paragraph{Average ball possession recovery time.} The average ball recovery time $\mbox{RecT}(g)$ is the average time elapsed between a team's last recorded pass and the first new pass made by a player of the same team. 

\paragraph{Shooting time.} The average shooting time $\mbox{ShotV}(g)$ is the average time between two shots of the same team. For example, in the men’s World Cup final, on average, for France approximately 345 seconds passed, and for Croatia about 281 seconds. 

\paragraph{Average pass length.} 
We measure the average pass length $\mbox{PassL}(g)$ in a match $g$ as the average Euclidean distance between a pass's starting and ending positions. 

\subsection*{Supplementary Information 2: PlayeRank scores}

The PlayeRank algorithm takes into account different types of events made by the players to compute the performance rating $r(u,g)$ of each player $u$ in a match $g$ \cite{playerank}. 
Given a match $g$, PlayeRank describes the performance of a player $u$ in $g$ by a n-dimensional feature vector $Q^g_u = [x_1,...,x_p]$, where each $x_j$, with $j=1,...,p$, is a feature describing a certain aspect of $u$'s behaviour during $g$. 
Some features are related to the number of specific events produced by $u$ in $g$ (e.g., passes, shots), others take into account the outcome of these events, e.g., whether or not they are accurate. The performance rating $r(u,g)$ of $u$ in $g$ is computed as:
\begin{equation}
{r(u,g)} = \frac{1}{R} \sum_{i=1}^p w_j x_j 
\label{ratingscore}
\end{equation}
where $w_j$ is the importance of feature $j$, $x_j$ the value of that feature, and $R$ a normalization constant. 
The weights $w_j$ are computed during a learning phase based on machine learning and consisting of two steps: feature weighting and role detector training \cite{playerank}. 
Note that PlayeRank assign every player to a role if they played at least $40\%$ of the matches in that role. 
Each role in the field is defined through a K-means clustering method implemented in the role detection phase of the learning phase \cite{playerank}. 
The performance rating $r(u, g)$ is combined with the number of goals scored using a goal weight $\alpha$ (set to $\alpha=0.10$ in our experiments). 
For example, Harry Kane (England), in the match against Panama, scored three goals and achieved a PlayeRank score of 0.59, demonstrating its centrality in the 6 to 1 victory. 
Similarly, the Australian champion Samantha Kerr, in the match against Jamaica, scored four times resulting in a PlayeRank score of 0.80.

\subsection*{Supplementary Information 3: Team Indicators}

\paragraph{H-indicator.} 
The H indicator summarizes different aspects of the passing behaviour of a team $T$ into a single value. 
All these aspects are related to the pass-based performance features, which are measured using a team's passing network in a certain match $g$. 
First, we compute the average amount $\mu_p$ of passes managed by players in a team during a match and the standard deviation $\sigma_p$ of the amount of passes managed by players in a team during a match \cite{Hind}. 
The higher $\sigma_p$, the higher is the heterogeneity in the volume of passes managed by the players. Moreover, we consider the distribution of passes over the zones of the pitch by splitting the football pitch into 100 zones, each of size 11 mt x 6.5 mt and computing the zone passing network, where nodes are zones of the pitch and edges represent the passes between two zones \cite{Hind}. 
We take the average amount $\mu_z$ of passes managed by zones of the pitch during the match and the standard deviation $\sigma_z$ of the amount of passes managed by zones of the pitch during the match \cite{Hind}. 
High values of $\sigma_z$ underlies the coexistence of hot zones with high passing activity and cold zones with low pass activity during the game. 
Low values of $\sigma_z$ indicates, however, a more uniform distribution of the pass in game activity across the zones of the pitch \cite{Hind}. 
Finally, we combine these indicators by their harmonic mean to summarize the passing behavior of a team $T$ into the H indicator: 
\begin{equation}
H(T,g) = \frac{5}{(1/w+1/\mu_p+1/\sigma_p+1/\mu_z+1/\sigma_z)}
\label{Hindicator}
\end{equation} 
\noindent where $w$ is simply the number of passes produced by the team $T$ in a match $g$.

\paragraph{Flow Centrality.} 
The team passing network allows measuring the centrality of each player within the network of passes. 
The team flow centrality derives from the player flow centrality \cite{amaral2010quantifying}, which we compute (and modify as needed) using the algorithm taken from \cite{pappalardo2019public}. 
The player flow centrality ranks each player based on their centrality in the network of passes in a certain match. 
Formally speaking, it measures the current-flow-betweenness-centrality value for each node (remembering that each node is a football player). 
The betweenness centrality captures a node's role in allowing information to pass from one part of the network to the other. 
Technically, it measures the percentage of shortest paths that must go through the specific node. 
The important thing to know is that betweenness is a measure of how important the node is to the flow of information through a network \cite{network}. 
In this context, it quantifies how central a player is in passing the ball from one side of the field to the other. 
The team flow centrality is then defined by setting on average the betweenness flow centrality values of players of the same team $T$ in the matches they played, $\mbox{FC}_{avg}(T,g)$. 
We also compute a function to measure the variability $\mbox{FC}_{std}(T,g)$ in the passing flow centrality of a team in a match. 
High values of $\mbox{FC}_{std}(T,g)$ highlight that there are players that individually are at the center of a team passing behavior in a particular game $g$; low values of $\mbox{FC}_{std}(T,g)$, otherwise, depict an equilibrium between players of the same team in the flow passing centrality.

\begin{figure}
    \centering
    \includegraphics[width=0.8\textwidth]{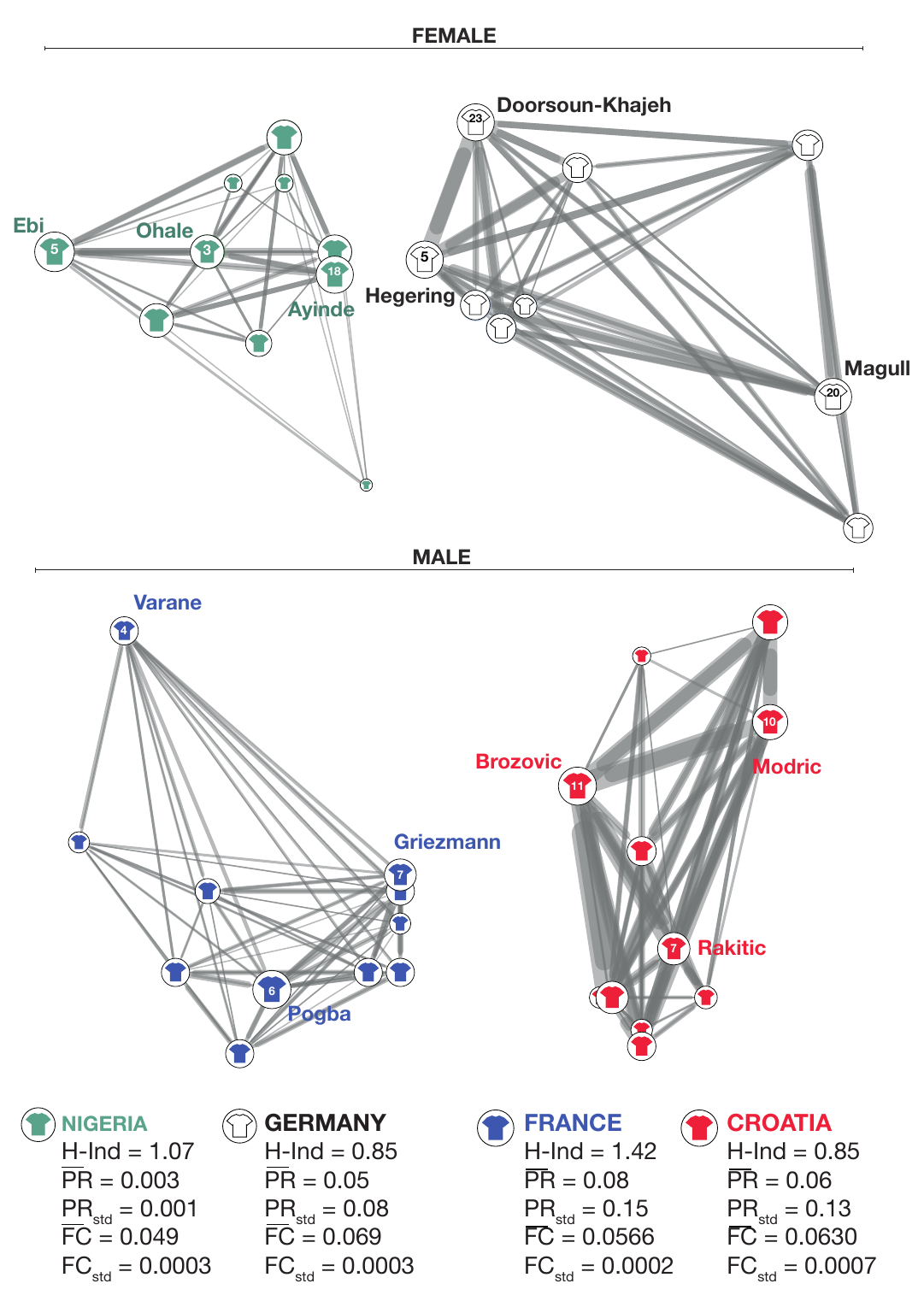}
    \caption{Passing network of the France World Cup game of the \textit{round of 16}, Germany v. Nigeria, and the Russia World Cup \textit{final}, France v. Croatia. Each node represents a player and its width is related to how many times teammates have passed the ball to that particular player. Formally, the width is related to the normalized weighted in-degree measure. The edges width, however, is weighted with respect to how many times two players have passed the ball to each other. There are highlighted the players who received the highest percentage of passes from their team mates, i.e., the most sought after on the pitch during the match. The algorithm used to draw the network was taken and modified as needed from the article \cite{pappalardo2019public}.}
    \label{networks}
\end{figure}

Supplementary Figure \ref{networks} shows two examples of passing networks and the corresponding H, FC, and PR values.

\subsection*{Supplementary Information 4: Predictions on Full Matches}

Tracing the classifier's predictions for those teams that competed in the same game can be interesting to verify whether the results and the comments made previously are not simply due to chance. To do this, we consider different \textit{test} sets (again based on the random state value with which the first \textit{training} set was divided); on each set we compute the class predictions, and we isolate the only games with both teams within the \textit{test} set. In particular, we use fifty different \textit{test} sets.

\begin{figure}
\centering
\subfigure[]{\includegraphics[scale=0.56]{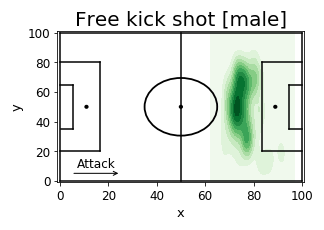}}
\subfigure[]{\includegraphics[scale=0.56]{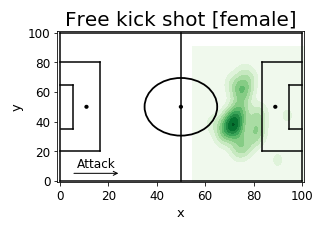}} \\
\subfigure[]{\includegraphics[scale=0.56]{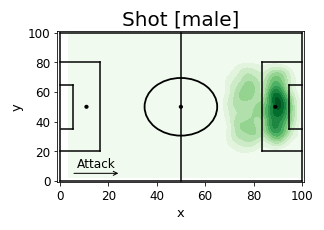}}
\subfigure[]{\includegraphics[scale=0.56]{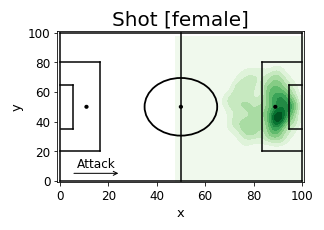}}
\caption{Heatmaps describing the pitch zones from where the \textit{free-kick shots} and the \textit{shots in motion} were more frequently made by male and female players, during their respective World Cup championships. They show the kernel estimate of the First Grade Intensity function $\lambda(s)$, where the event points $s_i$ are the \textit{free-kick shots} ((a) and (b)) and the \textit{shots in motion}((c) and (d)), and the football field is the region of interest $R$. The darker is the green, the higher is the number of \textit{free-kick shots} and \textit{shots in motion} in a specific field zone. The pitch length (x) and width (y) are in the range [0,100], which indicates the percentage of the field starting from the left corner of the attacking team.}
\label{heatmapS1}
\end{figure}

\begin{figure}
    \centering
    \includegraphics[width=0.8\textwidth]{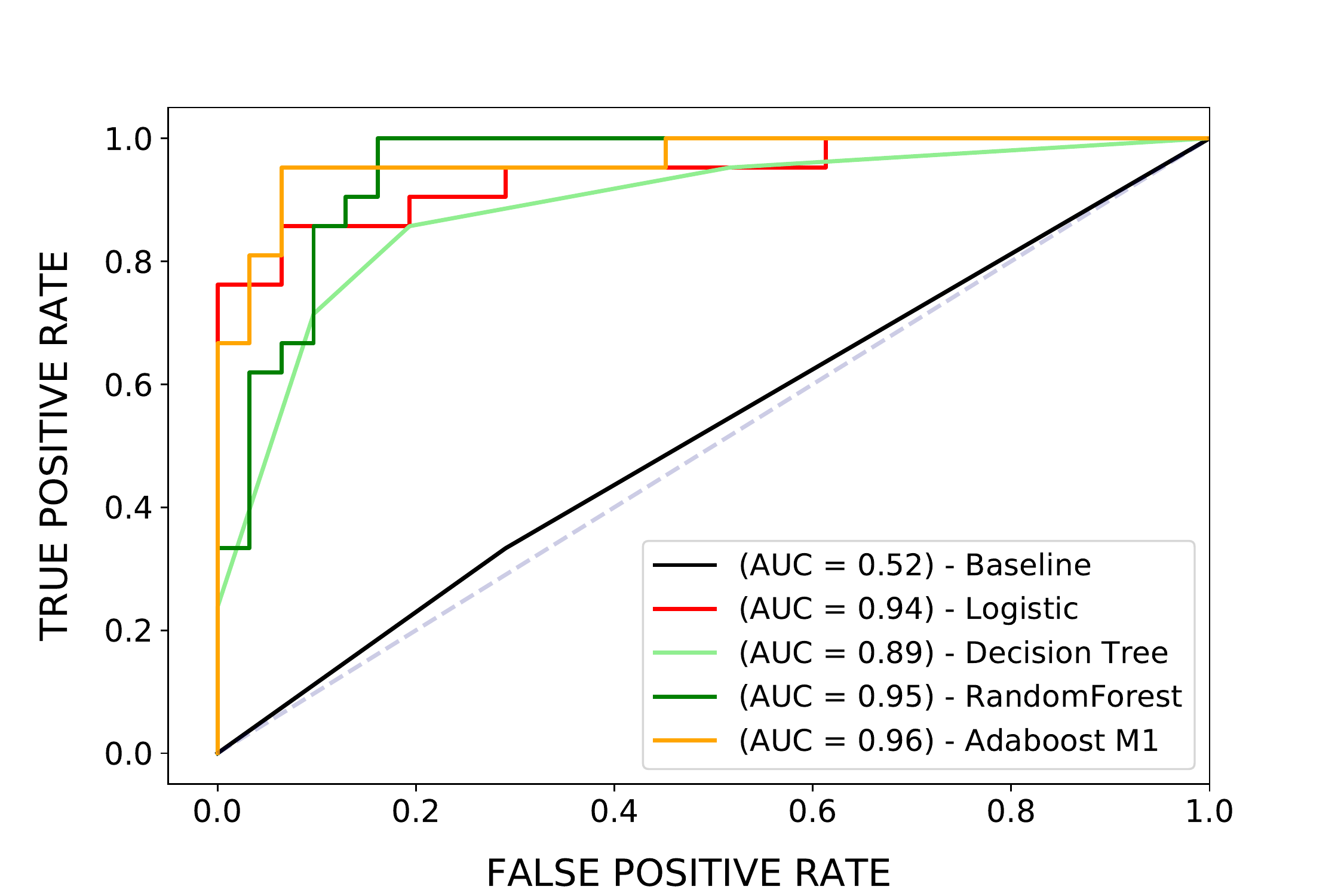}
    \caption{ROC curves for the implemented classifiers. They trace out the true positive rate and the false positive rate, as the probability threshold changes, i.e., the threshold beyond which an observation is assigned to class 1 (male team). 
    When the true positive rate and the false positive rate are both 0, the threshold is 1 (all the observations are classified as class 0) \cite[p. 147]{tibishirani}. 
    In this case, the true positive rate is the percentage of male teams correctly classified and the false positive rate is the percentage of female teams mistaken as male, using a given threshold. 
    The actual thresholds are not shown. 
    The AUC represents the area under the curve, the larger the AUC the better the classifier \cite[p. 147]{tibishirani}. 
    Random Forest and Adaboost M1 show the best predictive performance.} 
    \label{roc} 
\end{figure}

\begin{figure}
\centering
\subfigure{\includegraphics[width=0.45\textwidth]{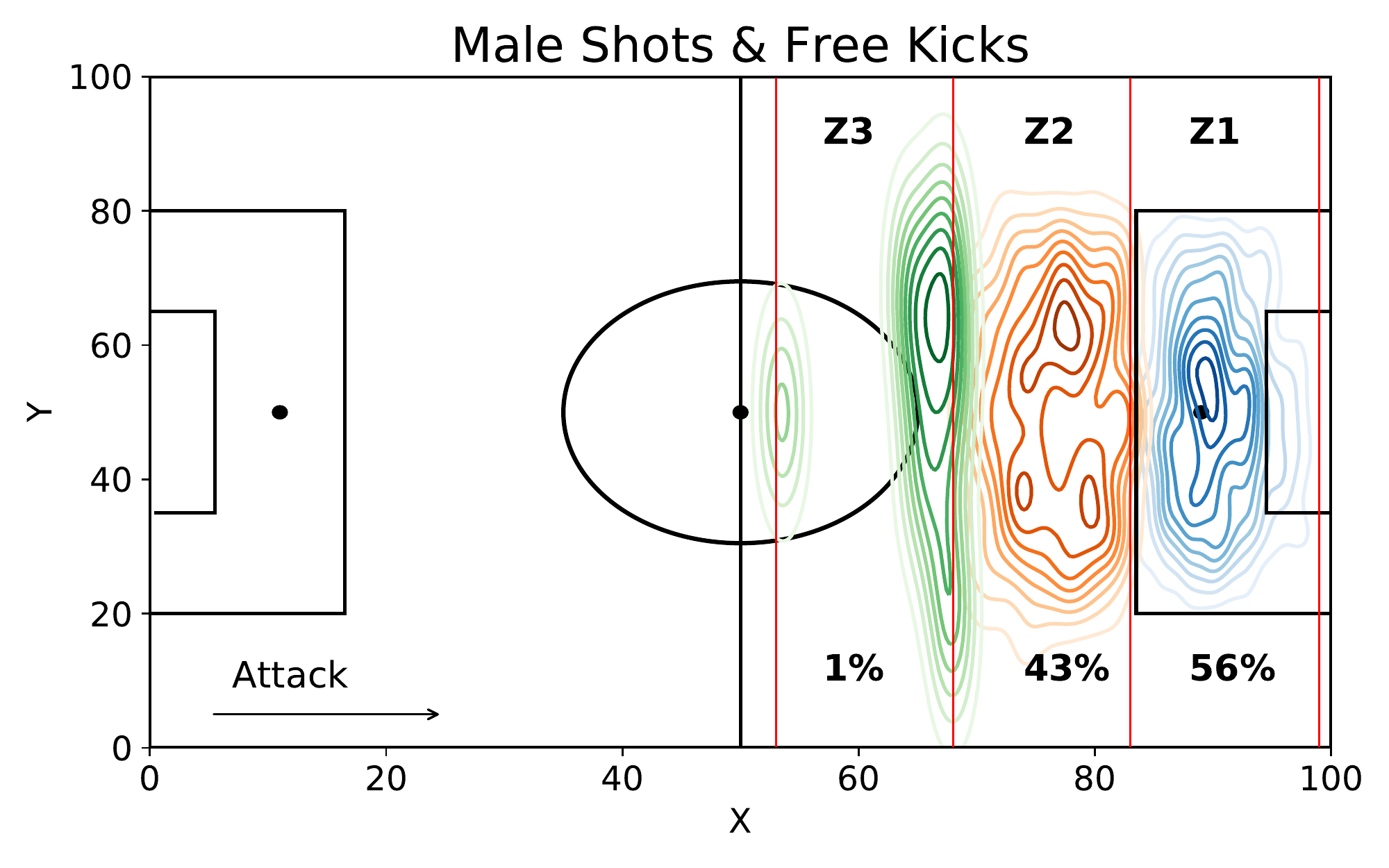}} 
\subfigure{\includegraphics[width=0.45\textwidth]{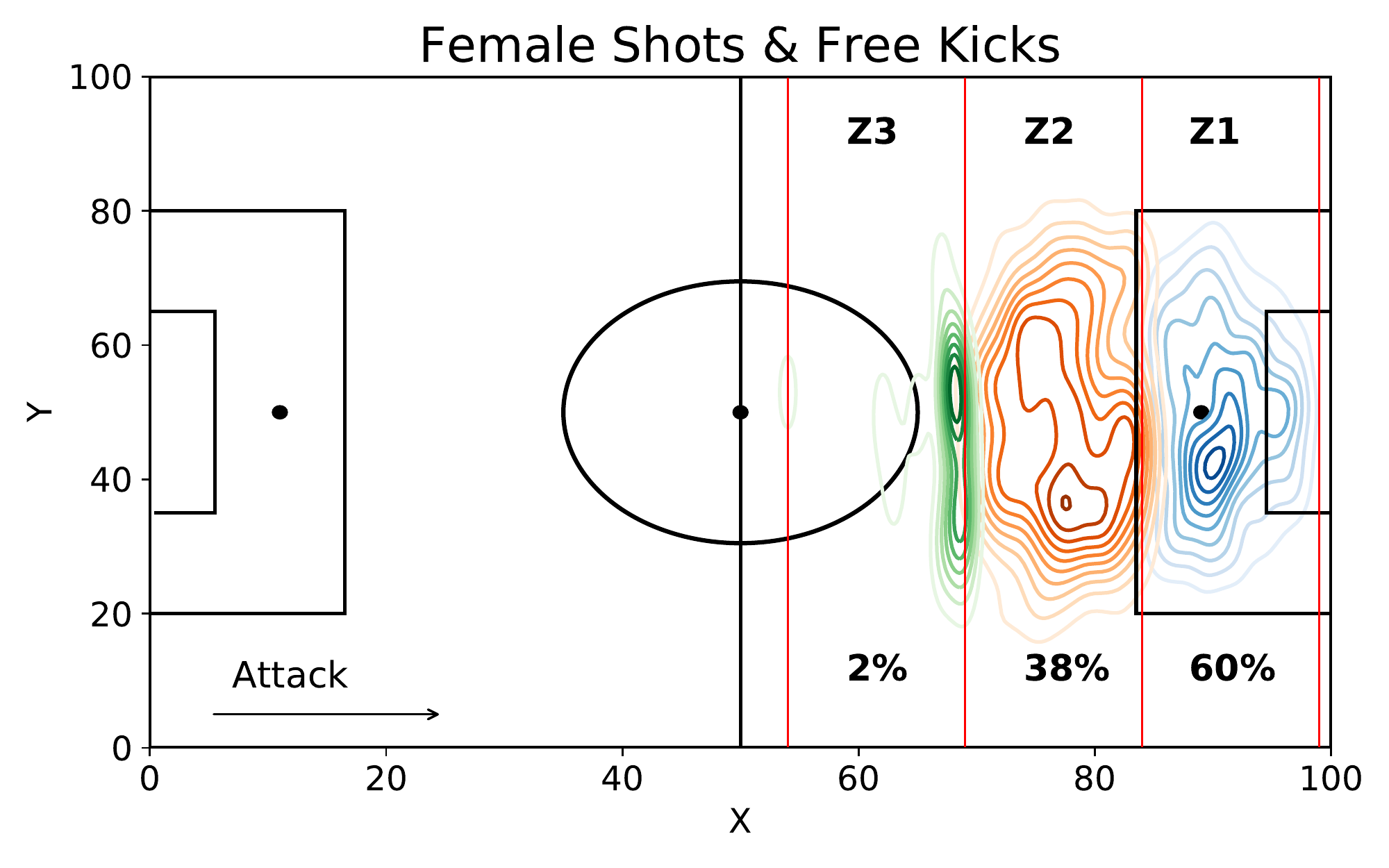}}
\caption{Pitch zones from where free-kicks and shots in motion are made by male players (a) and female players (b). 
We split the attacking midfield into three equal zones: $Z1$ is the area closest to the opponents' goal, $Z3$ the furthest and $Z2$ the zone in the middle. 
In each zone, we show the percentage of free kicks and shots in motion made in that zone, and depict the kernel estimate of the First Grade Intensity function $\hat\lambda(s)$, where the event points $s_i$ are the free-kicks and the shots in motion, and the football field is the region of interest $R$. 
The darker the color, the higher is the number of events in a specific field position.
Female zones are 1.1 meters closer to the opponents' goal than male zones. 
}
\label{shots&kicks}
\end{figure}

\section*{Acknowledgments}
We thank WyScout Spa for providing the match events, Daniele Fadda for his support on data visualization.

\section*{Funding}
This research has been supported by by EU project H2020 SoBigData++ RI, grant \#871042.

\section*{Authors' contributions}
LP directed the work, made statistical analysis, and wrote the paper. AR refined the statistical analysis and the classification experiments, made the plots. GP conducted the statistical analysis, the classification experiments, the plots, and wrote the paper. MN suggested experiments and checked the results. PC directed the work and wrote the paper.

\end{document}